\documentclass[]{aa}
\newcommand{\apropto}{\;
  \raise0.3ex\hbox{$\propto$\kern-0.75em\raise-1.1ex\hbox{$\sim$
  }}\;\hskip-2pt }
\usepackage[dvips]{graphicx}

\newcommand{\lta}{\;
  \raise0.3ex\hbox{$<$\kern-0.75em\raise-1.1ex\hbox{$\sim$
  }}\;\hskip-2pt }
\newcommand{\gta}{\;
  \raise0.3ex\hbox{$>$\kern-0.75em\raise-1.1ex\hbox{$\sim$
  }}\;\hskip-2pt }

\begin{document}
\title{Reversals of the solar dipole
}

   \author{D.~Moss\inst{1} \and L.\,L.~Kitchatinov\inst{2,3} \and D.~Sokoloff\inst{4}}

   \offprints{D.Moss}

   \institute{School of Mathematics, University of Manchester, Oxford Road,
 Manchester, M13 9PL, UK
 \and
 Institute for Solar-Terrestrial Physics, PO Box 291, Irkutsk, 664033, Russia
 \and
 Pulkovo Astronomical Observatory, St. Petersburg 196140, Russia
 \and
 Department of Physics, Moscow University, 119992 Moscow, Russia }

   \date{Received ..... ; accepted .....}

\abstract{During a solar magnetic field reversal the magnetic dipole moment does not vanish, but migrates between poles, in contradiction to
the predictions of mean-field dynamo theory.}{We try to explain this as a consequence of magnetic fluctuations.}{We exploit the statistics of fluctuations
to estimate observable signatures.}{Simple statistical estimates, taken with
results from mean-field dynamo theory, suggest that a non-zero dipole moment
may persist through a global field reversal.}
{Fluctuations in the solar magnetic field may play a key role in explaining reversals of the dipolar component of the field.}


\titlerunning{Reversals of the solar dipole}
\authorrunning{Moss et al.}

\maketitle

\section{Introduction}
\label{int}

Magnetic activity is a quasi-regular cyclic process during the
course of which the solar magnetic field changes its polarity in
each nominal 11-year cycle. This cyclic activity is usually
explained as a manifestation of solar dynamo action somewhere within
the solar interior. The Sun and its activity are more or less
axially symmetric. Of course, the distribution of sunspots is not
completely symmetric because of, e.g., the discrete nature of this
tracer of solar activity. However, if distributions of this and
other tracers of solar activity are averaged over a reasonable time
interval, they becomes almost axisymmetric, and  at any time surface deviations
from axisymmetry in the form of preferred longitudes are about 10\%
(e.g. Berdyugina et al. 2006). This is why a description of the
solar activity in terms of axisymmetric dynamo models seems to be a
reasonable step, and nonaxisymmetric features can be considered as
small perturbations. Of course, a deeper understanding of solar
activity in terms of direct numerical simulations  for detailed
solar magnetic configurations, which automatically include
small-scale nonaxisymmtric features is a very desirable subsequent
step  (see, e.g. Brown et al. 2010, whose model has a dipole moment
whose evolution does hint at some of the desired features).

The scheme described above looks plausible. However it implies that
the mean magnetic dipole moment of the magnetic field
vanishes during the course of each reversal. The point is that
axisymmetric spherical mean-field dynamos have two possible
directions of their dipole magnetic moment, parallel or
anti-parallel to the rotation axis. Mean-field solar dynamo models
exploit this idea massively. Almost all mean-field models assume
that the magnetic dipole moment vanishes at the instant of field
reversal.

However, on the other hand observers insist that in reality the
magnitude of the solar magnetic dipole moment is reduced at the times of its
reversal, but does not vanish exactly. Instead its direction moves
continuously from one pole to the other on a quite complicated
trajectory, which varies from one reversal to the other (e.g.
Livshits \& Obridko 2006). The topic was investigated in detail by
DeRosa et al. (2012) who provided convincing evidence concerning the
migration of the solar magnetic dipole moment from one pole to the
other during the course of a reversal.

Of course, it can be claimed that the solar dynamo is not fully
axisymmetric and that weak deviations from axial symmetry in solar
hydrodynamics and/or dynamo generated solar magnetic field are
sufficient to explain the relatively weak dipole magnetic moment at
the times of reversals. The point however is that the mean-field
solar dynamo is known to be very robust and it is extremely
difficult to excite a quite substantial nonaxisymmetric
magnetic field in a solar type mean-field dynamo.

A radically different viewpoint would be that
that there is no need to
describe solar magnetic field evolution in the framework of
mean-field  dynamos and that if such models have problems, it is time to
move to direct numerical simulations of the non-averaged induction
equation, e.g. along the lines of Brown et al. (2010).
We feel however that it would be much better
to attempt to
resolve the problem rather to ignore it.
Mean-field dynamo theory is too useful a tool in understanding these
phenomena to be lightly discarded.

We note that the reversals with nonvanishing magnetic moment touches on
one more closely related problem. The point is that an inclined
magnetic moment creates a quadrupole magnetic field mode in addition
to the nonaxisymmetric one. The appearance of a quadrupole component
was convincingly addressed in the context of interest by DeRosa et
al. (2012). We note however that dynamo excitation of a quadrupole
dynamo configuration by a spherical dynamo is much less problematic
(see e.g. Moss et al. 2008) 
than excitation of nonaxisymmetric configurations.

The aim of this paper is to give an interpretation of observations
discussed in DeRosa et al. (2012) in the context of nonaxisymmetric
dynamo configurations and spherical mean-field dynamos. We start by
revisiting the problem of the existence of nonaxisymmetric solutions in
the framework of mean-field dynamos, and confirm that it struggles
to explain the phenomenology under discussion. Then we suggest a
minimal extension of the mean-field theory which can explain the
phenomenon, by adding explicitly a random (fluctuating) magnetic
field component (which is anyway implicitly present in any
mean-field dynamo) to the dynamo generated mean field. We
demonstrate that such model can explain the nonaxisymmetric magnetic
dipole directions during the reversal time, within the admissible
parameter range. DeRosa et al. (2012) provide an accessible
background for the ideas developed in this paper.

\section{Searching for nonaxisymmetries in solar dynamos}
\label{sear}

The topic of symmetries of spherical dynamo was extensively
addressed in the early years of dynamo studies and the basic message
from that epoch still remains valid. In particular, R\"adler (1986)
and Moss et al. (1991a) investigated spherical dynamo models which
permit nonaxisymmetric solutions, and found that solar-type dynamos
preferentially excite axisymmetric rather than nonaxisymmetric
magnetic fields, and that the axisymmetric solutions are very stable
with respect to nonaxisymmetric perturbations. In fact, we did rerun
some similar models and confirmed the result. A point to be
mentioned in the present context is that the lifetime of a
nonaxisymmetric perturbation introduced in a dynamo generated
axisymmetric magnetic configuration can exceed the reversal time, i.e.
if the desired perturbation arises it can survive through the period
of reversal.

Of course, a dynamo can in some situations excite nonaxisymmetric
configurations and efforts were undertaken to include such
possibilities in the framework of solar dynamos (e.g. Ruzmaikin et
al. 1988; Moss 1999). Such attempts are rather isolated
exploratory studies and at the moment they do not provide strong
motivation for further reconsideration of significant
nonaxisymmetric field generation in more-or-less conventional
mean-field solar dynamos. We can note that, for example in Moss
(1999) the axisymmetric and nonaxisymmetric field components
oscillate in phase.

Note that the issue of possible excitation of a quadrupole
axisymmetric magnetic configuration is quite different from
that of the existence of nonaxisymmetric solutions. Of course, the solar dynamo is observed
normally to excite a dipolar configuration. However a modest
variation in the profiles of the dynamo drivers and/or control
parameters can result in excitation of a configuration with
quadrupolar symmetry; this problem was recently revisited by Moss et
al. (2008), among many others. 
Moreover, archival solar activity data for the XVIIIth
century (Arlt 2009) give a hint that the solar magnetic field had
quadrupolar symmetry at that time (Illarionov et al. 2011). A
substantial deviation from dipole symmetry with a significant
admixture of a quadrupole field is known to have been present at the
end of the Maunder Minimum (Sokoloff \& Nesme-Ribes 1994).
Nevertheless, the above remarks do not provide evidence for
quadrupole-like type features playing a role in solar magnetic field
reversals.

We do not give here a detailed verification of the above results, but
restrict ourselves to just one example.

%
%

For orientation, we performed a simple minded numerical experiment.
We took as a basis the nonlinear nonaxisymmetric dynamo code described in the
investigation of solar active longitudes by Berdyugina et al. (2006).
This code uses an approximation to the solar rotation law in the
"convection zone" proper (fractional radius $r\ge 0.7$), with an
isotropic alpha-effect. The model also includes a rather thick overshoot
layer with slightly reduced diffusivity, which allows the angular velocity
to become uniform at its base ($r=0.64$), in a smooth manner.

\begin{figure}
\begin{center}
\includegraphics[width=  8.5 cm]{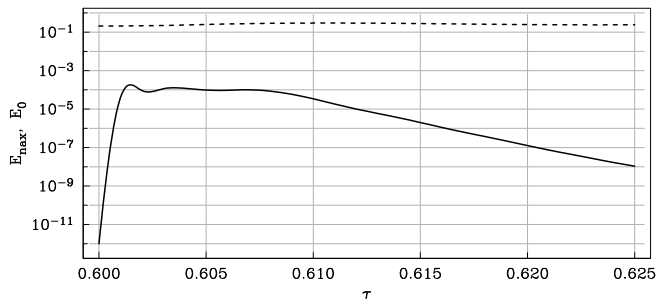}
\end{center}
\caption{\label{naxosc}
The evolution of energy in the axisymmetric field (upper curve) and
nonaxisymmetric field (lower curve). At dimensionless time 0.6, the steadily
oscillating axisymmetric solution (period $P_{\rm E}\approx 0.024$)
is perturbed by the addition of a nonaxisymmetric
part to alpha between times 0.6 and 0.605. The nonaxisymmetric perturbation
is then removed, and the nonaxisymmetric field decays.
}
\end{figure}

First the code was run with slightly supercritical parameters,
to obtain a stable, oscillating solution with
pure dipole-like parity. In our dimensionless units, the period of oscillation
in energy is $P_{\rm E}\approx 0.024$ (this would correspond to
the "11 yr" sunspot cycle on the Sun). Then a massive alpha-perturbation was
imposed in one longitudinal hemisphere, rotating with the angular
velocity at the base of the convection zone ($r=0.7$), for a time interval
of length 0.005 (i.e. about $0.2P_{\rm E}$). The perturbation was then
switched off.  The time evolution of the energies in the axisymmetric and
nonaxisymmetric (overwhelmingly in mode $m=1$) parts of the field are shown in
Fig.~\ref{naxosc}.

Fig.~\ref{naxosc} illustrates two relevant points.
A nonaxisymmetric perturbation
in alpha rapidly generates a (weak) nonaxisymmetric field.
When the driver is removed
(at time 0.605 in Fig.~\ref{naxosc}), the nonaxisymmetric field decays
with a decay time of about $25\%$ of the cycle period (i.e. of $P_{\rm E}$).
(Because of the disparate magnitudes of the quantities plotted in
Fig.~\ref{naxosc} the oscillation in the axisymmetric field is barely visible.)
We do not pretend that this model is in any way realistic, but it does
illustrate the points that a nonaxisymmetic field is quickly established
by the perturbation, and decays much more slowly.
This is quite consistent with previous studies which indicate that
axisymmetric solutions for $\alpha\Omega$ dynamos in spherical geometry are
very stable to nonaxisymmetric perturbations, e.g. R\"adler (1986), Moss et al. (1991b).

\section{A scenario for reversal: mean magnetic field plus fluctuations}
\label{sce}

A minimal extension of standard mean-field dynamo theory that
includes solar magnetic reversals with a nonvanishing magnetic
moment can be presented as follows. Mean-field dynamo theory assumes
that, apart from the mean magnetic field $\bf B$ which is considered
to be the large-scale magnetic field, magnetic field fluctuations
$\bf b$ are also present, so the total magnetic field ${\bf H} =
{\bf B} + {\bf b}$. The mathematical expectation of $\bf b$ is zero
but the spatial average of $\bf b$ remains finite. This is because
the number of convective cells $N$, while large, is not so big that
the spatial averaging of $\bf b$ (which scales as ${\bf b} / \sqrt
N$) effectively vanishes .

The number of convective cells $N$ participating in dynamo action in
the convection zone
can be estimated as follows. Taking for estimates the
supergranulation scale as 30 Mm, we find that there are about $10^4$
supergranules at the solar surface or about $(5 \sim 7) \times 10^4$
in the whole convection zone. However
supergranulation is a surface phenomenon whose spatial scale is
probably controlled by the depth of  the He$^{++}$ ionization region
(\cite{Nea81}). The characteristic scales of the deeper convection are
larger, so that $N \approx 10^4$ seems to be a reasonable estimate
for the number of convection cells in the Sun.

Knowing the magnetic field $\bf H$ we can estimate the dipole magnetic
moment $\bf m$ as

\begin{equation}
{\bf m} = {\rm const} \int {\bf H} d^3 x = \bar {\bf m} + a {\bf m}_1 b/ \sqrt N.
\label{rep}
\end{equation}
where $\bar {\bf m}$ is the magnetic moment of the mean field $\bf
B$, ${\bf m}_1$ is a random vector of unit length,  $b$ is the
r.m.s. value of the magnetic fluctuations and $a$ is a numerical
constant.

Two mechanisms producing fluctuating magnetic fields are known:
the small-scale dynamo and the wiggling of large-scale field lines by
turbulent convection. The observation-based upper limit on the
amplitude of the surface fields produced by the small-scale dynamo
in the Sun is only  about 3~G (\cite{S12}). The contribution of such small
fields to the global magnetic dipole moment of Eq.~(\ref{rep}) can be neglected
in view of the large number $N$ of convective cells. We therefore
consider the distortion of global field lines by convection as the
primary source of fluctuating fields.

Assume that $b$ is of order of the field strength of the toroidal
magnetic field $B_T$. Such an estimate looks plausible in the framework
of the standard ideas of mean-field dynamos and is supported by the
analysis of unipolar sunspot group statistics (Khlystova \&
Sokoloff 2010; Sokoloff \& Khlystova 2011). The poloidal (polar)
field $B_P$ of the Sun, which determines $\bar{\bf m}$, is
$\sim$1\,G (cf., e.g. \cite{H10}). The toroidal field near the base
of convection zone is believed to be at least 1000 times stronger.
Near the surface, however, the toroidal field is probably weaker and the
poloidal field inside the convective zone is probably stronger
(\cite{KO12}). Taking for estimates $B_P = 0.03 B_T$ and $\sqrt N =
100$ we find that the first term in the r.h.s. of Eq.~(\ref{rep}) to be
several times larger than the second. In other words, it is the
contribution of the mean magnetic field that determines the total solar
magnetic moment far from the instant of reversal. Of course, the
direction of $\bf m$ does not coincide with that of $\bar m$, i.e.
the direction of the rotation axis precisely, but the angle $\theta$
between $\bf m$ and rotation axis is quite small, $\tan \theta = 0.1
\sim 0.2$, i.e. $\theta \approx 5^\circ - 10^\circ$.

The situation near the time of magnetic field reversal is quite
different. Then $\bar {\bf m}$ vanishes and the total magnetic
moment $\bf m$ is determined by the second term in the r.h.s. of
Eq.~(\ref{rep}). $\bf m$ becomes weak but does not vanish. With the
above estimates, the magnetic moment at the instant of reversal
becomes several times weaker than its characteristic value
far from the reversal epochs.

When the Sun is still far from the instant of reversal, the first
term in the r.h.s. of Eq. (1) is larger than the second. If ${\bf
m}_1$ is directed to the North pole, the direction of the total magnetic
moment is aligned closely with this pole. For the same reason, $\bf m$ is
aligned closely to the opposite pole after the reversal,
when $\bar{\bf m}$ is again large.  The dipole's
track from one pole to the other is determined by the direction
${\bf m}_1$, i.e. by the nonaxisymmetric part of the total magnetic
field $\bf H$.  $\bf H$ is a physical field, 
and convection cannot destroy it immediately.
This follows from the results of Sect.\,II where we find that the
flow destroys a nonaxisymmetric field in a time of about 1-2 years,
and is why the motion from one pole to the other is quasiregular.
It is not fully regular just because the flow does not instantly
 destroy deviations from axial symmetry.
 
 In summary,
$m$ is determined by the sum of the dominant fluctuations (nonaxisymmetric)
and the mean field contribution (axisymmetric). The nonaxisymmetric fluctuations
individually decay on a timescale $t_{\rm D}\la t_r$, and their sum is
determined predominantly by the several most recent 
contributions.
As the mean field component weakens, changes sign on passing through
zero and then grows again, the sum $m$ moves from being directed outwards in one hemisphere to pointing out from the other,
on an irregular path essentially determined by the
dominant fluctuation.

The relative contribution of the second, fluctuating term in the
r.h.s. of Eq.~(\ref{rep}) can also be estimated as a ratio of
reversal time to the cycle length. Taking 1-2 years for the reversal
time $t_r$ and 11 years for the cycle length, we find again that the
fluctuating part of $\bf m$ is up to an order of magnitude smaller
than its mean part.

According to Sect.~II
a nonaxisymmetric feature, and in
particular ${\bf m}_1$, can survive for up to 1-2 years, i.e. the
time $t_r$ required for a reversal. On the other hand, the
convection turnover time $t^*$ estimated for the vortices of the
largest scale is 1-2 solar rotation (cf. \cite{S91}), i.e. a few
months -- substantially less than the reversal time. Indeed,
observers refer to the so-called 'fast' global changes on that time
scale (see e.g. Hoeksema, 2006). Magnetic fluctuations, being
determined by cumulative action of the velocity field, have a longer
memory time, $t_r$, than that of the convection itself ($t^*$);
however some rapid changes on the time scale $t^*$ have to be
expected.

The presence of two memory times, $t_r$ and $t^*$, can explain the
irregular, chaotic track of the magnetic moment during the reversals.
Observers (e.g. Livshits and Obridko 2006) stress this feature of
the reversals.

\section{Discussion}

We have demonstrated that a minimum extension of the standard
mean-field solar dynamo theory to allow for fluctuations is
sufficient to describe magnetic field reversals with nonvanishing
magnetic moment. With this in mind the main concepts of mean-field
dynamos can be retained to explain the phenomenology of the solar
cycle with reasonable accuracy; further development looks desirable.

The initial formulation of mean-field dynamos included magnetic
fluctuations at least implicitly. Hoyng (1987) stressed their role
explicitly and Choudhuri (1992), Moss et al. (1992)
 and Hoyng (1993) applied this idea
to explanations of solar cycle variability. The contemporary level of
knowledge of fluctuations in the dynamo governing parameter suggests
that they may provide a scenario explaining solar Grand Minima
(e.g. Moss et al. 2008, Usoskin et al. 2009) and the Waldmaier relations
for the time dependence of the solar cycle (Pipin et al. 2012).
We conclude that the idea seems
relevant to explaining magnetic reversals also.

Note that a similar problem is present in the geodynamo. The
geomagnetic dipole is currently inclined to the Earth's rotation
axis at an angle $\theta = 11^\circ$ and, according to the
paleomagnetic data (e.g. Laj et al., 1991) the geomagnetic field has
reversed from time to time in geological history, with  nonvanishing
magnetic dipole moment. It is reasonable to assume that, magnetic
fluctuations underlie this phenomenon also.

In summary, we feel that recognizing the importance of fluctuations
in the context of the solar dynamo mechanism may provide a crucial step
to explaining a range of non-periodic phenomena.

\begin{acknowledgements}
LLK thanks the Russian Foundation for Basic Research
(project 12-02-92691\underline{\ }Ind) and the Ministry of
Education and Science of the Russian Federation (contract
16.518.11.7065) for their support. DS was supported by RFBR under
grant 12-02-00170. We are grateful to V.Obridko and V.Pavlov for
fruitful discussions.
\end{acknowledgements}

\end{document}